\documentclass[runningheads]{llncs}
\usepackage{amssymb}

\usepackage{dsfont}
\usepackage{bbm}

\usepackage{graphicx}
\usepackage{xcolor}
\usepackage{colortbl}

\usepackage[utf8]{inputenc}
\usepackage[final]{pdfpages}

\definecolor{LightGray}{rgb}{0.91,0.91,0.91}
\begin{document}
%
\title{
Linking Physicians to 
Medical Research Results 
via Knowledge Graph Embeddings and Twitter}

\titlerunning{Linking Physicians to 
Medical Research Results via KG Embeddings}
\author{Afshin Sadeghi\inst{1,2} \and
Jens Lehmann \inst{1,2}}
\authorrunning{A. Sadeghi and J. Lehmann}
%
\institute{Smart Data Analytics Group (SDA), University of Bonn, Germany \and
Fraunhofer Institute for Intelligent Analysis and Information Systems, Sankt Augustin \& Dresden, Germany\\
\email{\{sadeghi,jens.lehmann\}@cs.uni-bonn.de}\\
\email{\{afshin.sadeghi,jens.lehmann\}@iais.fraunhofer.de}}
\maketitle          


\begin{abstract}
Informing professionals about the latest research results in their field is a 
particularly important task
in the field of health care, 
since any development in this field directly improves the health status of the patients. Meanwhile, social media is an infrastructure that allows public instant sharing of information, thus it has recently become popular in medical applications. In this study, we apply Multiple Distance Knowledge Graph Embeddings (MDE) to link physicians and surgeons to the latest medical breakthroughs that are shared as the research results on Twitter. Our study shows that using this method physicians can be informed about the new findings in their field given that they have an account dedicated to their profession.
\keywords{Knowledge Graph Embeddings \and Social Media \and Social Good \and Health care  \and Twitter \and Machine Learning}
\end{abstract}
\section{Introduction}
Twitter is a projection of the interactions of a society connected to the internet, which is in constant evolution. The dynamic aspect of this social media allows manifold applications. From the rise of social media, Twitter was used to measure campaign impacts, collect opinions, analyze trends and to study crisis. However, recently, its applications are more individualized. Particularly, because Twitter has become the most popular form of social media used for health care communication~\cite{raghupathi2014big}, and it is reshaping health care~\cite{hawn2009take}, it has become the center of many studies in the field of health care. For example, a study suggests Twitter for knowledge exchange in academic medicine~\cite{twitterasatool} and it was argued that disease-specific hashtags and the creation of Twitter medical communities~\cite{pemmaraju2017disease} has improved the uniformity of medical discussions. Another study is dedicated to the influence of specific medical hashtags on social media platforms~\cite{brady2017colorectalsurgery}.

\textbf{Problem Statement:} Pershad et al.~\cite{pershad2018social} point out the potential of Twitter to reshape public health efforts, including disseminating health updates, sharing information about diseases. Especially, they emphasize on the role of Twitter to make research advances more accessible for physicians. They argue that connecting researchers and clinicians is crucial and useful since clinicians can use new information they discover from this closer contact with researchers to guide decision-making about patient treatments in such a field that is in constant progress.

\begin{figure}[h]
\label{Ontologyschema}
\caption{A schema of the medical professional knowledge graph on Twitter with example user instances. An orange dot line depicts a new link suggested by the proposed method.}
\centering
\includegraphics[width=12cm, height=9cm]{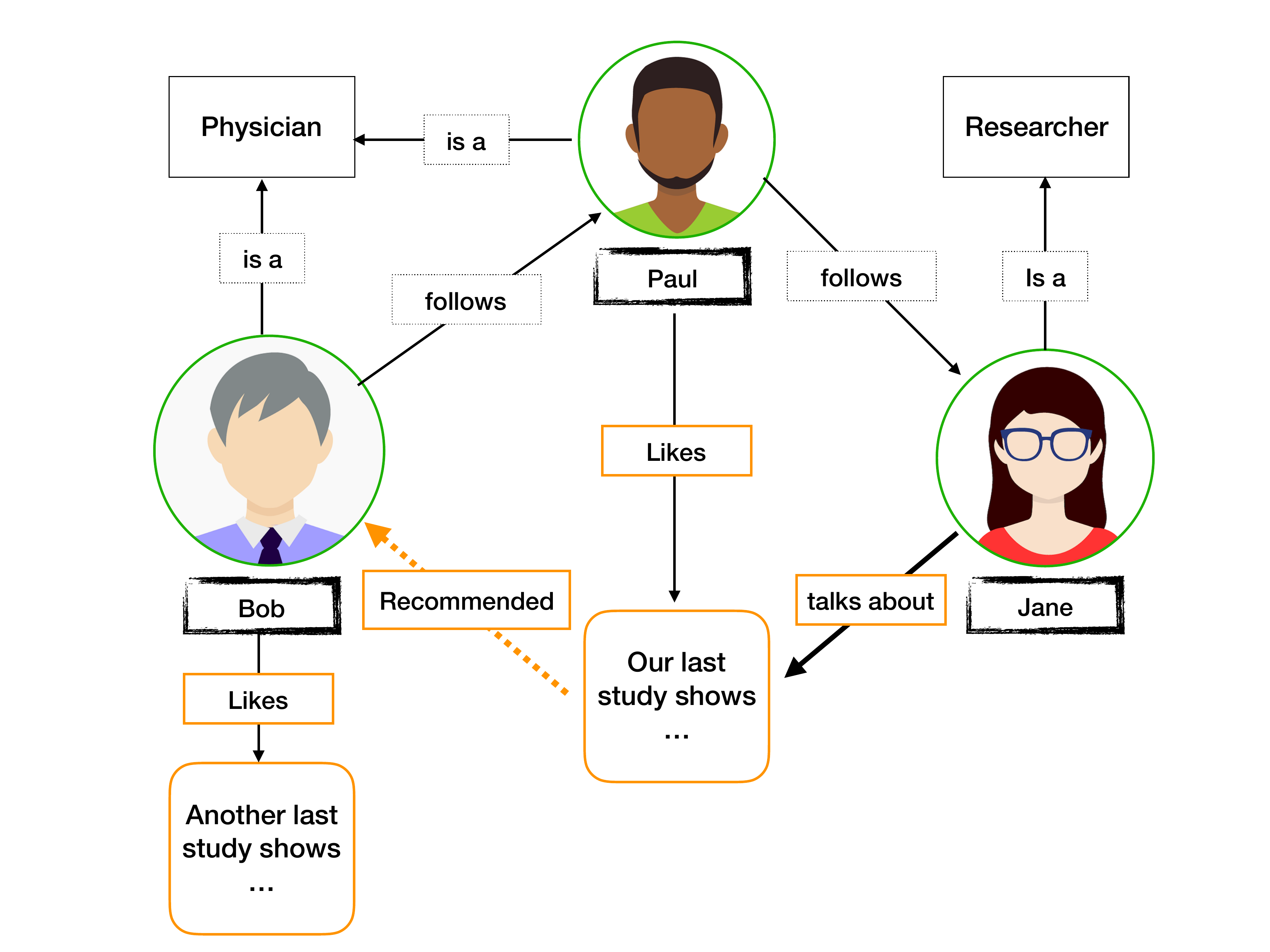}
\end{figure}

In this study, we target this problem by providing a method that suggests physicians and clinicians the recent research breakthroughs in their specialized field based on their current social activity. As the first step to reach this goal, we extract a subset of Twitter network and we generate a knowledge graph(KG) from the extracted data. Figure~\ref{Ontologyschema} depicts a schema of the KG with example user instances and the relations between them. In this figure, it is shown that our method recommends a Tweet of Jane, who is a researcher about her latest findings to Bob who is a surgeon. The method calculates a probability that such a Tweet will be useful to Bob based on his previous favored Tweets and the relation to other physicians that work in the same field. 

We then apply an embedding method to predict links that are likely to be serving to the physicians. The proposed application is different from the user recommendation service of Twitter~\cite{Gupta2013WFS} which recommends
users to follow or the works that discover similar users~\cite{goel2013discovering}. Here, we focus our study and evaluation on suggesting related Tweets.

The main contribution of this study is the application of knowledge graph embeddings for a new field, the extraction of a KG from Twitter for the proof of concept and experiments targeted at physicians to help them stay up-to-date in their field. 
 
\section{Background} 

In a social network, a node of the graph represents a person while edges that link the nodes correspond to relationships between people. The edges are also called "connections", "links". Examples of social networks are graphs that describe Facebook and Twitter. Link prediction, in general, is the task of predicting whether a link exists between a given pair of nodes or not.
\\

\textbf{Definition 1:} a knowledge graph is defined by (E, P, T), a set of entities e $\in$ E, a set of predicates p $\in$ P and a set of triples t $\in$ T. A triple $(e_i, p_k, e_j)$ is made of two entities and a predicate that connects them. 

In a KG, two entities can be connected by several predicates. When describing a social network by a KG, nodes are translated to entities and links are translated to predicates, however, in KG an ontology usually specifies the class of describes what types of entities and predicates can construct a triple. A relational learning model usually learns the relations of a KG. Particularly, embedding models are a class of relational learning models that produce vector representations of the entities and predicates and predict the missing links.
\\

\textbf{Definition 2:} link prediction in a KG means to predict the existence of a triple, i.e., whether a relation exists from two entities $e_i$ and $e_j$ and a $k$-th predicate.

\section{Related Work}
Classic link prediction methods on social media use graph properties of the social network or NLP feature of nodes to predict links between entities. For example, ~\cite{MartincicIpsic2017LinkPO} is base solely on graph features and \cite{almansoori2012link} uses a similar technique for the social networks in healthcare. Meanwhile,~\cite{adafre2005discovering} uses common words to cluster and rank nodes and based on that predicts the closely-ranked nodes to be connected. Another Study ~\cite{al2006link} uses a combination of graph features and keyword matches to train classifiers(SVM, Naive Bayes, etc) to predict if a link exists between two nodes.
\\

Most of the studies on link prediction of social networks focus on the problem of link existence. Where some methods attempt to find link weights and the number of links between the nodes~\cite{vartak2008survey}. An advantage of link prediction using KG embedding is that the type of links are also predicted since these KG embedding models distinguish the type of links. 

TransE~\cite{bordes2013translating} is an embedding model that is popular because of its simplicity and efficiency. It represents the entities in a KG by a relation between the vectors representing them. The score function describing these vectors in TransE is:

\begin{equation}\label{eq:0}
Score_{TransE} = \parallel h_i + r_i - t_i \parallel_p~
\end{equation}
where $n$ refers to $L_1$ or $L_2$ norm and $h_i$ and $t_i$ are the vector representations of an entity and $ r_i$ is the vector representations of a predicate. For training, TransE uses margin ranking loss as the loss function. The following Section describes embedding a KG extracted from Twitter using MDE model.

\section{KG Embeddings for Twitter Link Prediction}

Knowledge graph embedding models usually generate a prediction based on their score function. Nickel et al.~\cite{nickel2011three} suggests performing link prediction by comparing the score of a triple with some given threshold $\theta$ or by ranking the entries according to their likelihood that the link in question exists. We similarly use the Multiple-Distance Embedding (MDE) model~\cite{sadeghi2019mde}. In comparison to TransE, this model can learn several relational patterns and thus more can more accurately learn the hidden relation between the entities. Specifically, MDE can learn relations with symmetry, antisymmetry, transitive, inversion and composition patterns. The score function of this model is as follows:

\begin{equation}\label{eq:1}
Score_{MDE} = w_1 \parallel h_i + r_i - t_i \parallel_p~+~ w_2 \parallel h_j + t_j - r_j \parallel_p~+~ w_3 \parallel t_k + r_k - h_k \parallel_p - \psi
\end{equation}

where $\psi \in \mathbbm{R^+}$ is a positive constant. The loss function of this model is:

\begin{equation}\label{eq:2}
loss =  \beta_1 \sum_{\tau\in \mathds{T}^+} [f(\tau)- \gamma_1]_+ + \beta_2 \sum_{\tau'\in \mathds{T}^-} [\gamma_2 - f(\tau')]_+ 
\end{equation}

where $\gamma_1 , \gamma_2$ are small positive values and $\delta_0, \delta'_0 = 0$. $\beta_1, \beta_2 >0$ are constraints. 
The given loss minimizes the score of the positive samples. Therefore, the smaller the score of a triple, that relation is more probable. Base on this property of the loss function, we define a measure to estimate the existence of a predicate such that the more probable triples are given a higher score. 

We designate the division of the maximum score of a triple in the training set to the score of a triple A as the probability of the existence of A:

\begin{equation}\label{eq:3}
P_a = \frac{max(Score_{training-triples})}{Score_A }
\end{equation}

This definition is based upon the assumption that after the training, the model accurately predicates the triples of the training set.

The equation compares only the triples of the same type (with the same predicate). Thus, in predicting the triples for linking physicians to medical Tweets, we consider only triples with $like\_research\_Tweet\_id$ predicates.

\fboxsep=2mm
\fboxrule=1pt

\begin{figure}[ht!]
\label{tweet_sample}
\caption{A sample of the extracted Tweets about the recent medical studies. Each row shows the content of one of the extracted Tweets.}
\centering
\fcolorbox{black}{white}
{\includegraphics[width=11.5cm, height=4.3cm]{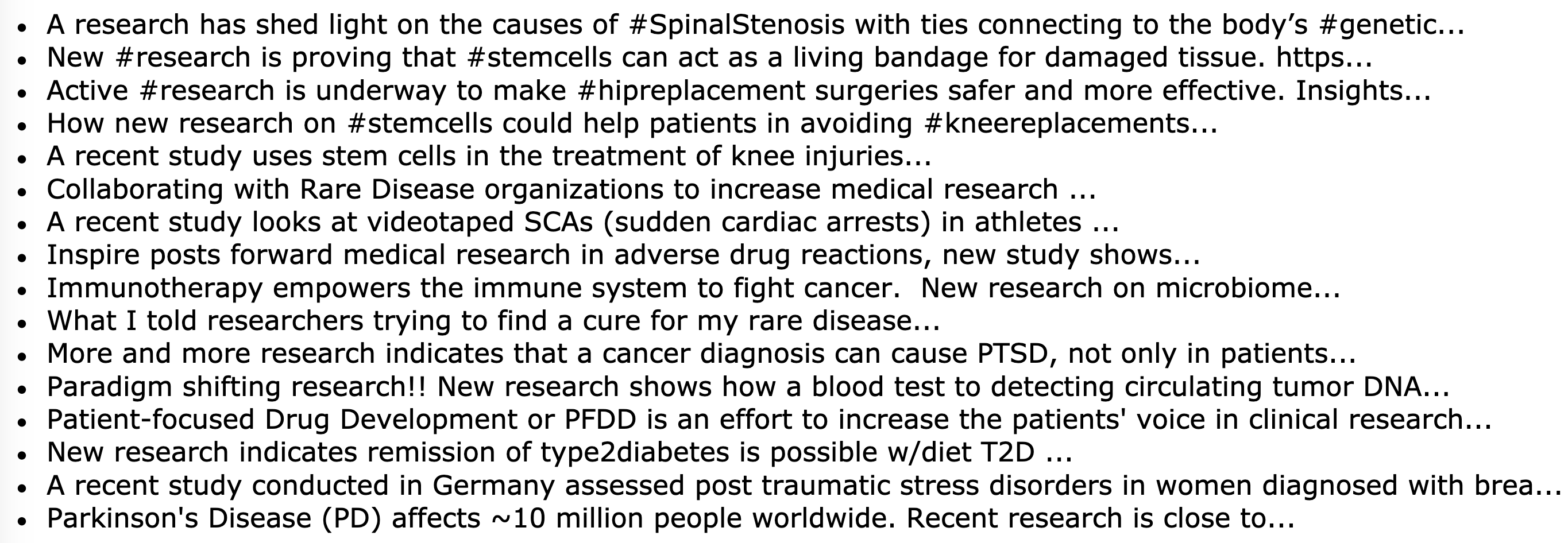}}
\end{figure}

To perform link prediction on Twitter, we train MDE over an extracted KG. In the following, we explain the procedure to extract the dataset from which we later generate a KG using it. 
\\

\textbf{Knowledge Graph Extraction:} We extract a set of Tweets about the latest medical studies using Python scripting and the Tweepy library~\footnote{\url{https://www.tweepy.org/}}. We filter our search by medical keywords and time in order to only obtain medical research related Tweets which were created from the beginning of the year 2019. Figure~\ref{tweet_sample} shows a sample of the extracted Tweets. To keep the privacy of users, we removed the user and Tweet identifiers from the figure.

With the same tools, we search for Twitter users who are physicians, surgeons, nurses, and researchers in the medical fields which have written about these topics or favored such Tweets. Our continues inquiry which took 8 hours, provided us 5996 Twitter users. Between these users, the job title of 69 instances was deductible (researcher in the medical field or physician) based on the medical job titles in their profile descriptions. This step reduced the users to 69 instances. We then extract the relations among these users and the relations among the users and the gathered Tweets. We also extract the users who follow or are followed by these users so we gain the neighbors of these users in the social network.

We then generate a multi-relational knowledge graph from scraped data by converting these relations to triples. To create these triples, we first define an ontology for the social KG. This ontology includes five types of relations. Table~\ref{tab:relTypes} lists these relation types. We also anticipate two classes for users in the ontology. Table~\ref{tab:userTypes} presents these classes. The created knowledge graph (TW52) includes 4439 entities which are comprised of 1021 users and 3418 Tweets. 
The final constructed KG includes 4791 triples. The anonymised dataset is openly available for research purposes in \url{https://git.io/fj6h8}. 
 \begin{table}[!ht]
    \centering
\begin{minipage}[t]{0.49\linewidth}\centering
\caption{Relation types in the Social Ontology}
\label{tab:relTypes}
    \begin{tabular}{c|c}
        Relation Id &   Relation \\
        \hline
        0 & \texttt{is\_talking\_about} \\
        \rowcolor{LightGray}1 & \texttt{is\_followed\_by} \\
        2 & \texttt{is\_following}  \\
        \rowcolor{LightGray}3 & \texttt{job\_title\_type\_is} \\
        4 & \texttt{likes\_research\_Tweet\_id} \\
    \end{tabular}
\end{minipage}\hfill
\begin{minipage}[t]{0.49\linewidth}\centering
\caption{Class of users in the Social Ontology}
\label{tab:userTypes}
    \begin{tabular}{c|c}
            Class Id &  User Entity Class \\
        \hline
0 & \texttt{job\_title\_medical\_researcher} \\
\rowcolor{LightGray}1 & \texttt{job\_title\_physician} \\
\end{tabular}
\end{minipage}
\end{table}


\section{Experiments}
We set up two experiments. We firstly evaluate how well the MDE method performs on the social media dataset against a baseline in the task of link prediction. We then analyze the suggestion results of the model in different situations.

\subsection{Performance Evaluation}
We set up an experiment to evaluate the link prediction performance of MDE against TransE as the baseline. 
\\

\textbf{Evaluation Setup:}
We dedicate 80 percent of the knowledge graph extracted from Twitter as the training dataset and set the rest as the test dataset. We randomly choose triples by uniform random selection to separate them for the test set. We perform ranking the score of each test triple against its versions with replaced head, and once with a replaced tail. We then compute the hit at N(hit@N), mean rank(MR) and mean reciprocal rank (MRR) of these rankings. We set the vector size of 
TransE to 20 and choose the vector size of 10 for MDE. We use $L_2$ normalization to normalize their score function and train them by 700 iterations. For MDE, we set the hyperparameters as follows: $\gamma_1 = \gamma_2 =3$ and $\psi = 1.2$.
\\

\textbf{Results:}
Table~\ref{tab:ranking} lists the evaluation results of TransE and MDE on the extracted knowledge graph. Due to the sparsity of the graph, TransE gains very low ranking scores while MDE produces superior results for all the MR and MRR and hit@N tests. The results suggest the positive influence of relation patterns learning in MDE. 
\\

\begin{table}[h]
\centering
\resizebox{0.60\linewidth}{!}{
    \begin{tabular}{c|c c c c c}
        Model & MR & MRR & Hit@1 & Hit@3 & Hit@10\\
        \hline
    \rowcolor{LightGray}TransE & 1327 & 0.021 & 0.005 & 0.019 & 0.048 \\
    MDE & \textbf{1287} & \textbf{0.148} & \textbf{0.071} & \textbf{0.161} & \textbf{0.332} \\
    \end{tabular}
}
\caption{Results on Twitter extracted dataset (TW52). Better results are in bold.}
\label{tab:ranking}
\end{table}

\subsection{Link Prediction Analysis:}
In this section rather than studying the performance of the model, we establish an experiment to analyze the suggestion results of the model to find out whether it creates sound suggestions in different situations. We apply the model to learn on the constructed KG and then we use it to suggest the possible interesting research results for the physicians suggested. We then study the suggested results.
\\

Considering the physicians in the KG and Tweets which include research results, we calculate the probability that such a Tweet is favorable for physicians using Equation~\ref{eq:3}. In our experiment, the hit@1 of the training triples was 99.8 percent, therefore, assuming maximum probability for the training triples in the formula holds for the experiment.

The observation of relations and entities in the KG shows that it is structured with the small world network patterns~\cite{watts1998collective}. Particularly, it includes hub users and Tweets which are connected to other nodes with a number of links that greatly exceed the average degree in the network.

We select a subset of physicians in the KG and classify them according to their relation to other users and Tweets into 4 groups of 5 users. We particularly inspected their relation to hub users, which we call them User type A. Users of type A are followed by a large number of users(at least 200), they are active users and have favored variant Tweets. We consider also Users of type B which follow a small number of users(25) who are also physicians or researchers. Table~\ref{tab:results} lists these groups of users and the mean of their probability to like a Tweet C that includes a research study. We consider two Tweets similar if their representative vectors have a small angle. These Tweets are usually favored by the same group of people.

\begin{table}[!ht]
    \centering
\caption{Mean Probability of linking to a Tweet C for  users with different communities and liked Tweets}
\label{tab:results}
    \begin{tabular}{c|c}
        User group &  Mean Probability of C\\
        \hline
        Users U that follow A. A and U like a Tweet similar to C  & 0.205 \\
        \rowcolor{LightGray}Users U that follow A. A likes a Tweet similar to C &  0.134\\
        Users U that follow B. A and B like a Tweet similar to C & 0.975  \\
        \rowcolor{LightGray}New users U that still follow nobody and like no Tweet & 0.127 \\
    \end{tabular}
\end{table}
It is observable from Table~\ref{tab:results} that in the proposed model, users that follow a diverse group of users and topics, are less likely to be interested in an inquired Tweet than those with less diverse connections. This effect is even stronger than when the user has liked a similar Tweet before. This suggests that the model performs better if a Twitter account is dedicated only to social communications related to her profession.

Additionally, the new users that have not favored any Tweet are expected the least among the users to favor a Tweet. 

\section{Discussion of the specificity of the problem}
The proposed experiment in the study has two major components. The first is data extraction and KG construction section which we specify it to the problem by data cleaning and filtering the extracted data and creating an ontology specific to the physicians and research related tweets. The result of this part is TW52 knowledge graph which is sparse in comparison to the conventional benchmark datasets of embedding models, i.e., WordNet18 and FB15K.

The second part of the study is the MDE embedding model. Although MDE is a general method for the link prediction problems, the evaluations showed that it is capable of embedding the sparse dataset much better than the state-of-the-art TransE model. Therefore we consider both components appropriate for the proposed problem.

\section{Conclusion}
We proposed the usage of multiple-distance knowledge graph embeddings (MDE) to suggest Tweets about medical breakthroughs to physicians. We extracted a KG of medical research Tweets and their relations to the users which medical researcher or physicians.

We evaluated MDE against TransE as the baseline in a link prediction test for the social network KG. Our experiment shows the superior ranking performance of MDE over the baseline. 
We defined a probability for link suggestion and provided an analytic study for it. We thereby conclude that the model can be suggested to serve in connecting the physicians and the up-to-date advances in the medical studies. Considering the time constraints of physicians on social media~\cite{campbell2016social}, automating such suggestions can help physicians to find news and trends relevant for medical research results more easily and in less time.

As future work, it would be interesting to extend this study on a large scale and provide it as a live service. In addition, future studies may investigate the social effect of such application to find its effect benefits for patients besides the physicians.

\section{Acknowledgements}
This study is partially supported by project MLwin (Maschinelles Lernen mit Wissensgraphen, grant no. 01IS18050F)\footnote{\url{https://mlwin.de/}} and Cleopatra (grant no. 812997). The authors gratefully acknowledge financial support from the Federal Ministry of Education and Research of Germany (BMBF) which is funding MLwin and European Union Marie Curie ITN that funds Cleopatra, as well as Fraunhofer IAIS.

\bibliographystyle{splncs04}

\bibliography{references}
\end{document}